\documentclass{llncs}
\usepackage{graphicx,times,hyperref}

\begin{document}

\title{swMATH -- a new information service for~mathematical~software\thanks{The final publication is available at http://link.springer.com.}}
\author{Sebastian B\"{o}nisch\inst{1} \and Michael Brickenstein\inst{2} \and Hagen Chrapary\inst{1} \and Gert-Martin Greuel\inst{3} \and Wolfram Sperber\inst{1}}
\authorrunning{Sebastian B\"{o}nisch et al}
\tocauthor{Sebastian B\"{o}nisch, Michael Brickenstein, Hagen Chrapary,Gert-Martin Greuel, and Wolfram Sperber}
\institute{FIZ Karlsruhe/Zentralblatt MATH, Franklinstr. 11, 10587 Berlin, Germany
\and
Mathematisches Institut Oberwolfach, Schwarzwaldstr. 9-11, 77709 Oberwolfach-Walke,Germany. 
\and
Technische Universit\"{a}t Kaiserslautern, Fachbereich Mathematik, Postfach 3049,\\ 67653 Kaiserslautern, Germany}

\maketitle

\begin{abstract}
An information service for mathematical software is presented. Publications and software are two closely connected facets of mathematical knowledge. This relation can be used to identify mathematical software and find relevant information about it. The approach and the state of the art of the information service are described here. 
\end{abstract}

\section{Introduction}
In 1868, the first autonomous reviewing journal for publications in mathematics -- the ``Jahrbuch \"{u}ber die Fortschritte der Mathematik'' --  was started, a reaction of the mathematical community to the increasing number of mathematical publications. The new information service should inform the mathematicians about recent developments in mathematics in a compact form. Today, we encounter a similar situation with mathematical software. Until now, a comprehensive information service for mathematical software is still missing. We describe an approach towards a novel kind of information service for mathematical software. A core feature of our approach is the idea of systematically connecting mathematical software and relevant publications.

\section {The state of the art}
  
There have already been some activities towards the development of mathematical software information services. A far-reaching concept for a semantic web service for mathematical software was developed within the MONET project \cite{MONET} which tries to analyze the specific needs of a user, search for the best software solution and organize the solution by providing a web service. However, the realization of such an ambitious concept requires a lot of resources.
Also, a number of specialized online portals and libraries for mathematical software were developed. One of the most important portals for mathematical software is the Netlib \cite{Netlib} provided by NIST. Netlib provides not only metadata for a software but also hosts the software. Netlib has developed an own classification scheme, the GAMS \cite{GAMS} system, which allows for browsing in the Netlib.  Other important manually maintained  portals, e.g., ORMS \cite{ORMS}, Plato \cite{Plato} or the mathematical part of Freecode \cite{Freecode}, provide only metadata about software. 

\section{The publication-based approach}     
\vspace*{-\baselineskip}
Mathematical software and publications are closely interconnected. Often, ideas and algorithms are first presented in publications and later implemented in software packages. On the other hand,  the use of software can also inspire new research and lead to new mathematical results. Moreover, a lot of publications in applied mathematics use software to solve problems numerically. The use of the publications which reference a certain software is a central building block of our approach.
\begin{description}
\item[Identification of software references in the zbMATH database]
There are essentially two different types of publications which refer to a software, publications describing a certain software in detail, and publications in which a certain software is used to obtain or to illustrate a new mathematical result. 
In a first step, the titles of publications were analyzed to identify the names of mathematical software. Heuristic methods were developed to search for characteristic patterns in the article titles, e.g., `software', `package', `solver' in combination with artificial or capitalized words. It was possible to detect more than 5,000 different mathematical software packages which were then evaluated manually. 
\item[Software references in publications -- indirect information of software]
The automatically extracted list of software names (see above) can be used as a starting point for searching software references in the abstracts: More than 40,000 publications referring to previously identified software packages were found in the zbMATH database. Of course, the number of articles referring to a given software is very different, ranging from thousands of publications for the 'big players' (e.g. Mathematica, Matlab, Maple) to single citations for small, specialized software packages.
An evaluation of the metadata of the publications, especially their keywords and MSC classifications has shown that most of the information is also relevant for the cited software and can therefore be used to describe the latter. For instance, we collect the keywords of all articles referring to a certain software and present them in swMATH as a keyword cloud, common publications and the MSC are used to detect similar software.
\item[More information about software]
Web sites of a software -- if existing -- are an important source for direct information about a software. As mentioned above, there are also special online portals which sometimes provide further information about certain mathematical software packages.
\item[Metadata scheme for software]
The formal description of software can be very complex. There are some standard metadata fields which are also used for publications, like authors, summary, key phrases, or classification. 
For software however, further metadata are relevant, especially the URL of the homepage of a software package, version, license terms, technical parameters, e.g. , programming languages, operating systems, required machine capacities, etc., dependencies to other software packages (some software is an extension of another software), or granularity.
Unfortunately, often a lot of this metadata information is not available or can only be found with big manual effort. The focus of the metadata in swMATH is therefore on a short description of the software package, key phrases, and classification. For classification, we use the MSC2010 \cite{MSC2010} scheme even though the Mathematics Subjects Classification is not optimal for software.
\item[Quality filter for software]
swMATH aims at listing high-quality mathematical software. Up to now, no peer-reviewing control system for software is established. However, the references to software in the database zbMATH can be used as an indirect criterion for the quality of a software: The fact that a software package is referred in a peer-reviewed article also implies a certain quality of the software.
\end{description}

\section {Further software}
\vspace*{-\baselineskip}
There are several reasons which suggest an extension of the publication-based approach. A major drawback of the publication-based approach is the time-delay between the release of software and the publication of an article describing the software. This delay can be up to several years for peer-reviewed journals.
A second reason, not every software is referenced in peer-reviewed publications. Often, software is described in technical reports or conference proceedings.

Also, not all publications describing or using mathematical software are contained in the zbMATH database, e.g., if a software was developed for a special application and articles on it were published in a journal outside the scope of Zentralblatt MATH.

In order to build a comprehensive information service about mathematical software, we therefore still use other sources of information as online portals for mathematical software, contacts to renowned mathematical institutions, research in Google and other search engines with heuristic methods. One problem here is the quality control of this software. Being listed on a renowned portal for mathematical software should be a clear indicator for the quality of a software, whereas a mere Google hit does not mean much with respect to quality.

\section {Sustainability}
swMATH is a free open-access information service for the community. The development and maintenance of it, however, are not for free. For sustainability, the resources needed for the maintenance of the service must be minimized. Automatic methods and tools are under development to search for mathematical software in the zbMATH database, and to maintain and update the information on software (e.g.\ an automatic homepage verification tool).

In order to ease the maintenance of the service, the developments of the user interface and the retrieval functionalities are carried out in close coordination with the corresponding developments in zbMATH. The swMATH service enhances the existing information services provided by FIZ Karlsruhe/Zentralbatt MATH. The integration of the database swMATH in the information services of Zentralblatt MATH contributes to its sustainability. At the moment, links from software-relevant articles to zbMATH are provided. In the near future, back links from zbMATH to swMATH will be added too.

\section {The swMATH prototype}
The first prototype of the swMATH service was published in autumn 2012. Currently, the service contains information about nearly 5,000 mathematical software packages. It can be found at \url{http://www.swmath.org}.
The user interface of swMATH concentrates on the essentials, containing simple search and an advanced search mask.
Then a list of the relevant software is presented.
\begin{figure}[p]
    \centering
    \includegraphics[width=\textwidth]{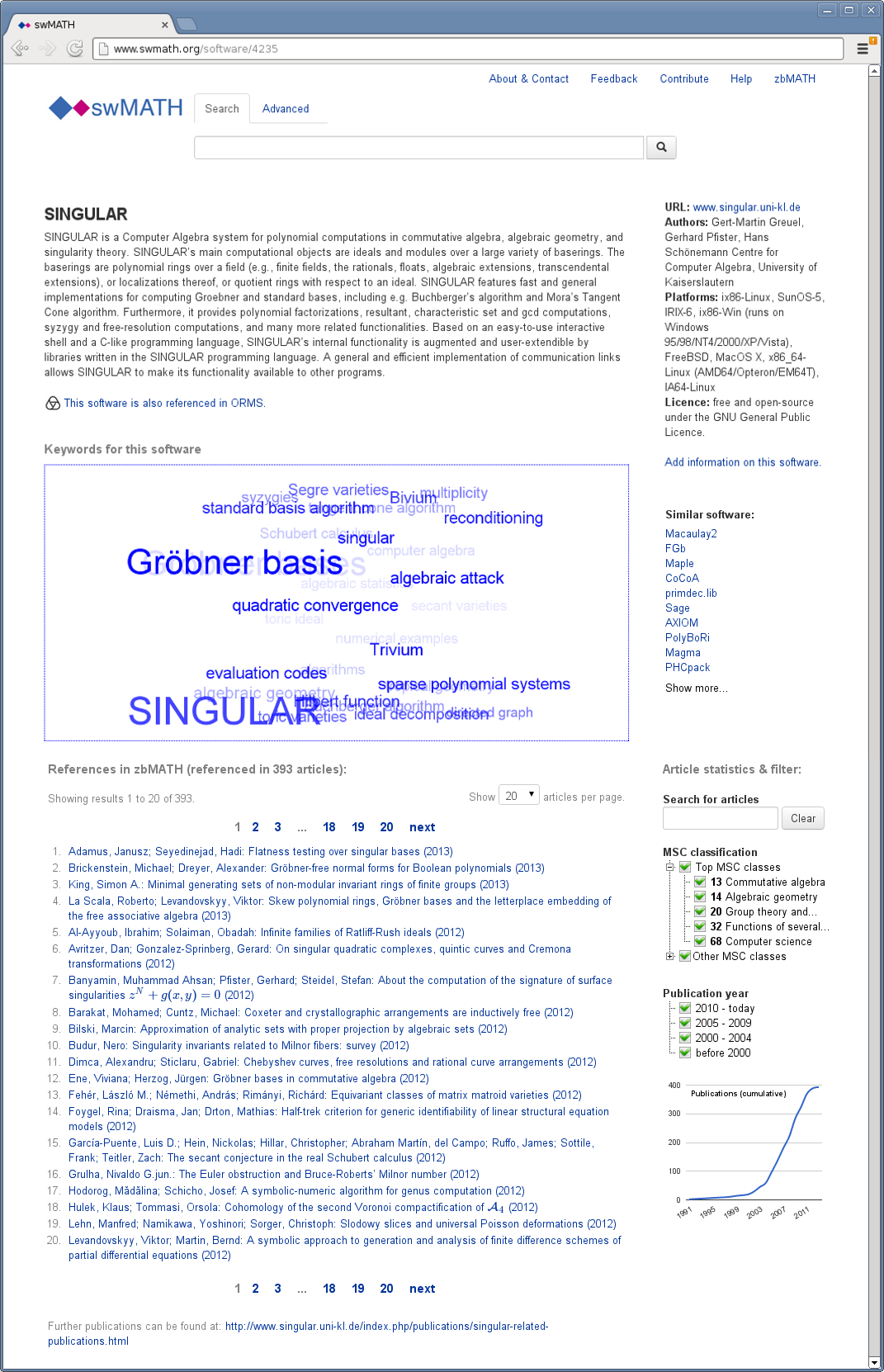}
    \caption{The detailed information for the software ''Singular''}
    \label{detailedpresentation}
\end{figure}

\noindent The detailed information about this software is shown if the name is clicked. It contains  a description of the software, a cloud representation of key phrases (auto-generated from the key phrases of the publications), the publications referring to the software, the most important MSC sections, similar software and a plot showing the number of references over time. The latter is  an indicator for usefulness, popularity and acceptance of a package within the mathematical community.

\section{swMATH -- an information service under development}
swMATH is a novel information service on mathematical software basing on the analysis of mathematical publications. Automatic tools periodically check the availability of URLs. Further heuristic methods to automatically extract relevant information from software websites are currently developed.
Another possibility to keep the software metadata up-to-date is direct contact with (selected) software authors and providers.

So far, the software identifiers in swMATH are not persistent. However, for the productive release of swMATH persistent identifiers are planned.

The user interface is under permanent development; we recently added a browsing feature and will further enhance the usability of the swMATH web application. In order to meet the demands of the mathematical software community, we created an online questionnaire which has recently been distributed to several thousand participants, \url{https://de.surveymonkey.com/s/swMATH-survey}.

We hope that the swMATH service will be a useful and broadly accepted information service for the mathematical community.

\end{document}